\documentclass[twocolumn,showpacs,superscriptaddress,prl,aps,floatfix]{revtex4}

\usepackage{graphicx}
\usepackage{rotating}
\usepackage{amsmath}

\newcommand{\br}{{\bf r}}

\newcommand{\brp}{{\bf r'}}

\begin{document}

\title{Open shells in reduced-density-matrix-functional theory}

\author{N.\,N.\,Lathiotakis} 
\affiliation{Institut f{\"u}r Theoretische Physik, Freie Universit{\"a}t Berlin, Arnimallee 14, D-14195 Berlin, Germany}

\author{N.\,Helbig}
\affiliation{Institut f{\"u}r Theoretische Physik, Freie Universit{\"a}t Berlin, Arnimallee 14, 
D-14195 Berlin, Germany}

\author{E.\,K.\,U.~Gross}
\affiliation{Institut f{\"u}r Theoretische Physik, Freie Universit{\"a}t Berlin, Arnimallee 14, 
D-14195 Berlin, Germany}

\begin{abstract}
Reduced-density-matrix-functional theory is applied to open-shell systems. We introduce 
a spin-restricted formulation by appropriately expressing approximate correlation-energy functionals
in terms of spin-dependent occupation numbers and spin-independent natural orbitals.
We demonstrate that the additional constraint of total-spin conservation is indispensable for the proper
treatment of open-shell systems.
The formalism is applied to the first-row open-shell atoms. The obtained ground-state
energies are in very good agreement with the exact values as well as other state of the art quantum chemistry calculations.
\end{abstract}
\maketitle
The proper description of electronic 
correlation is the central problem in theoretical material science. Density functional theory (DFT) deals 
with this problem by considering the electronic density as the fundamental variable. 
DFT is built upon the Hohenberg-Kohn theorem\cite{HK} which ensures that
the expectation value of every observable, including the total energy, is a unique functional of the 
electronic ground-state density.
The many-electron problem is then mapped onto an auxiliary, non-interacting system, 
the so-called Kohn-Sham system\cite{KS} whose density reproduces the interacting density. In practice,
the so-called exchange and correlation part of the Kohn-Sham potential needs to be approximated. 
DFT is tremendously successful in describing properties of real materials and that success 
is proved by its general acceptance as one of the major tools in exploring 
theoretically the world of real materials. Nevertheless, there are certain materials, termed collectively
as highly correlated materials, for which the results of DFT, at least with standard functionals
for exchange and correlation, deviate significantly from the experimental values. 

Reduced-density-matrix-functional theory (RDMFT) is an alternative way to deal with the many-electron problem. 
It is based on Gilbert's theorem\cite{gilbert}, which guarantees that the expectation value 
of any observable of the system in the ground-state is a unique
functional of the one-body reduced-density-matrix (1-RDM) 
\begin{equation}
\gamma(\br,\brp)\! =\! N \!\!\! \int \!\!d^3\br_2 \! \cdots\! d^3\br_N \Psi^*(\brp, \br_2\! \cdots\! \br_N ) 
\Psi(\br, \br_2 \!\cdots\! \br_N ) 
\end{equation}
where $\Psi$ is the many-body ground-state wavefunction.  The advantage of this approach, 
compared to DFT, is that the exact many-body kinetic energy is easily expressed in terms
of $\gamma$. 
Although the properties of 1-RDM functionals had been the subject of theoretical studies for a long 
time\cite{gilbert,valone,zumbach,mueller}, there were few practical applications of RDMFT until recently. 

We emphasize that it is not possible to construct a Kohn-Sham-like independent electron scheme in RDMFT:
Due to its non-idempotency, the 1-RDM of an interacting system cannot be reproduced by a non-interacting
system because the latter always has an idempotent 1-RDM. This is reflected in the eigenvalues
of the 1-RDM which are equal to zero or one for a non-interacting system, while in the interacting
case some of them are fractional.

The first implicit functional of the 1-RDM was introduced in 1984 by M\"uller\cite{mueller}.
This functional was later derived from modeling the 
correlation hole, by Buijse and Baerends\cite{baerends}. Goedecker and Umrigar\cite{UG}
introduced a self-interaction correction to this functional. Other functionals have been proposed more
recently\cite{csanyi,csgoe,yasuda,kollmar,kios1,kios2}.
Implicit 1-RDM functionals depend explicitly on the so-called natural orbitals $\varphi_a$
and the corresponding occupation numbers $n_a$
which are defined as the eigenfunctions and the eigenvalues of the 1-RDM
\begin{equation}
\int d\brp^3\: \gamma(\br, \brp) \: \varphi_a (\brp) = n_a \: \varphi_a (\br)\,.
\end{equation}
The total energy can be minimized with respect to the natural orbitals and the occupation numbers, 
instead of the 1-RDM itself, by considering the appropriate subsidiary conditions, namely the conservation of the 
total number of electrons, the $N$-representability constraint and the orbital orthonormality.
Goedecker and Umrigar\cite{UG}  calculated the 
correlation energy of small atomic systems including the open-shell Lithium and Carbon atom.
Despite their simplicity, 1-RDM functionals are able to provide a good 
approximation of the correlation energy of small 
systems\cite{baerends,UG,csanyi,csgoe,yasuda,kollmar,staroverov,herbert}. 
Generally, the correlation energy of finite systems is improved by removing the self-interaction
terms\cite{staroverov,herbert}. However the correct dissociation limit of the 
H$_2$ molecule is found only if self-interaction is retained\cite{baerends,staroverov,herbert}.

In the present work, we demonstrate how natural orbital functionals in RDMFT 
can be applied to open-shell systems. For that 
purpose, we employ the M\"uller functional~\cite{mueller} with the self-interaction
correction proposed by Goedecker and Umrigar\cite{UG}. Our formulation is
spin-restricted, i.e. we have spin-dependent occupation numbers but spin-independent natural
orbitals. The advantage of this approach is that particular spin configurations are 
prescribed, in the same manner as in the restricted open-shell Hartree-Fock (ROHF) method\cite{roothaan}.  
We discuss the necessity of a spin-dependent constraint for the conservation of both the total number of 
electrons as well as the spin. 

We start with the 1-RDM functional described in Refs.~\cite{mueller,baerends,UG}, where we explicitly include 
the full spin dependence in the occupation numbers and the natural orbitals. We refer to this as the
spin-unrestricted functional that reads
\begin{multline}
\label{eq:ug}
E=\sum_{a,\sigma} n_a^{\sigma} h_{aa}^{\sigma} 
+ \frac{1}{2} \sum_{ab,\sigma\sigma^\prime}
n_a^{\sigma} n_b^{\sigma^\prime} (1-\delta_{ab}\delta_{\sigma\sigma^\prime}) 
J_{ab}^{\sigma\sigma^\prime} \\
-\frac{1}{2} \sum_{ab,\sigma}
\sqrt{n_a^{\sigma} n_b^{\sigma}} (1-\delta_{ab}) 
K_{ab}^{\sigma}\,,
\end{multline}
with $n_a^{\sigma}$ being the spin-dependent occupation numbers and 
$h_{aa}^{\sigma}$, $J_{ab}^{\sigma\sigma^\prime}$ and $K_{ab}^{\sigma}$ the one- and two-electron
integrals for the natural orbitals $\varphi_{a\sigma}$ 
\begin{eqnarray}
h_{aa}^{(\sigma)} & = & \int\! d^3\br \: \varphi_{a\sigma}^*(\br) \left[ -\frac{\nabla^2}{2} + 
V_{ext}(\br) \right] \varphi_{a\sigma}(\br)\,, \nonumber \\ 
J_{ab}^{\sigma\sigma^\prime} & = & \int\!\!\!\int\!\!\! d^3\br\: d^3\brp 
\frac{\varphi_{a\sigma}^*(\br) \: \varphi_{a\sigma}(\br) \: \varphi_{b\sigma^\prime}(\brp)\:
\varphi_{b\sigma^\prime}^*(\brp)}{| \br - \brp |}\,, \label{eq:ints}\\
K_{ab}^{\sigma} & = & \int\!\!\!\int\!\!\! d^3\br\: d^3\brp 
\frac{\varphi_{a\sigma}^*(\br) \: \varphi_{b\sigma}(\br) \: \varphi_{a\sigma}(\brp)\: 
\varphi_{b\sigma}^*(\brp)}{| \br - \brp |}\,. \nonumber
\end{eqnarray}
In the above expression, $V_{ext}(\br)$ is the external potential,
i.e. the ionic potential for atomic and molecular systems. The Kronecker $\delta$s 
are inserted in Eq.~(\ref{eq:ug}) in order to exclude explicitly the self-interaction terms\cite{UG}. 

We now concentrate on the spin-restricted case by assuming spin-independent
orbitals but still spin-dependent occupation numbers. Then the 
expression~(\ref{eq:ug}) reduces to
\begin{multline}
E= \sum_a \: (n_a^{\uparrow} + n_a^{\downarrow}) \: h_{aa}  + \\ 
\frac{1}{2} \sum_{ab} \left[ \left(n_a^{\uparrow} n_b^{\uparrow} + 
n_a^{\downarrow} n_b^{\downarrow}\right)
\left(1-\delta_{ab}\right) 
+ \left(n_a^{\uparrow} n_b^{\downarrow} + n_a^{\downarrow} n_b^{\uparrow}
\right) \right] J_{ab} \\
-\frac{1}{2} \sum_{ab} \left[ \left(\sqrt{n_a^{\uparrow} n_b^{\uparrow}} + 
 \sqrt{n_a^{\downarrow} n_b^{\downarrow}}\right) \left(1-\delta_{ab}\right) \right] K_{ab}\,. \label{eq:ourfunc}
\end{multline}
In this equation, the integrals $h_{aa}$, $J_{ab}$ and $K_{ab}$ are analogous to
those in Eq.~(\ref{eq:ints}) but for spin-independent orbitals $\varphi_a$.
This is the functional we employ in this work. It can be viewed as a generalization
of the ROHF functional in the sense that it reduces to the ROHF functional if all the occupation numbers 
are either zero or one. 
Although the formulation in this paper refers to the specific functional given in Eq.~(\ref{eq:ourfunc}),
the generalization to different functionals of similar kind\cite{csanyi} is easily achieved by
replacing the square roots in the exchange-like term 
in Eq.~(\ref{eq:ourfunc}) by the appropriate functions 
$f(n_a^{\uparrow}, n_b^{\uparrow})$, $f(n_a^{\downarrow}, n_b^{\downarrow})$.

Driven by physical requirements,
Cioslowski et al\cite{ciospernalziesce} have derived a list of criteria 
that 1-RDM functionals should fulfill. In addition, they examined whether
functionals of the M\"uller-type satisfy these criteria. 
Our open-shell version~(\ref{eq:ourfunc}) of the self-interaction-corrected M\"uller functional
satisfies the same criteria as the original M\"uller functional\cite{mueller} with the 
self-interaction correction of Goedecker and Umrigar\cite{UG}.

We now discuss the subsidiary conditions that have to be enforced in the minimization procedure 
of functionals like the one given by Eq.~(\ref{eq:ourfunc}).
Since the occupation numbers are spin-dependent we face a dilemma concerning the
conservation of the total spin. More specifically we have the following two options:
(i) to use one constraint for the conservation of the total number of electrons. This introduces a 
single Lagrange multiplier, the chemical potential $\mu$. In that way, we unfortunately allow for
charge transfer from one spin channel to the other, usually from the majority to the minority spin and the
total spin is not preserved, or (ii) to use an extra constraint for the conservation of the total spin, making
the minimization more restrictive. In practice, we use two different constraints for the
spin-up and the spin-down electrons. In that way, we introduce two Lagrange multipliers, or, in other words,
a spin-dependent chemical potential. It is one of the goals of the present work to assess these 
two different ways of minimizing the energy functional. Of course, the above dilemma applies only to 
open-shell systems. It is expected that the first option, being less restrictive, leads to a lower total
energy. However, since RDMFT (like DFT) is not variational, a lower energy is not necessarily better.
Hence, it is not a priori clear whether enforcing 
the spin conservation constraint in addition to the particle number conservation will improve or worsen the 
energy. 

Additionally, as in the case of closed-shell systems, we have to include two further
subsidiary conditions. The first is the orbital orthonormality. Unfortunately, all 
1-RDM functionals are not invariant under unitary transformation of the natural orbitals.
This leads to a complex minimization problem 
which consists of orbital dependent Fock-like operators and 
off-diagonal Lagrange multipliers. The equations we have to solve in order to
find the orbital solution for fixed but fractional occupation numbers have
the form
\begin{eqnarray}
F^{(a)}(\br) \varphi_a(\br) & = & \sum_b \epsilon_{ab} \: \varphi_b(\br)\,, \nonumber \\
\epsilon_{ba} & = & \epsilon_{ab}\,, 
\end{eqnarray}
where $F^{(a)}$ is analogous to the Fock matrix in Hartree-Fock 
theory but in this case orbital-index dependent, and  $\epsilon_{ab} $ are the
non-diagonal Lagrange multipliers, which should be Hermitian according to the second equation.
The orbital-index dependent operator $F^{(a)}$ is 
\begin{multline}
F^{(a)}(\br)= (n_a^{\uparrow} + n_a^{\downarrow}) \: h(\br) + \\ 
\sum_{b} \left[ \left(n_a^{\uparrow} n_b^{\uparrow} + 
n_a^{\downarrow} n_b^{\downarrow}\right)
\left(1-\delta_{ab}\right) 
+ \left(n_a^{\uparrow} n_b^{\downarrow} + n_a^{\downarrow} n_b^{\uparrow}
\right) \right] {\cal J}_{b}(\br) \\
- \sum_{ab} \left[ \left(\sqrt{n_a^{\uparrow} n_b^{\uparrow}} + 
 \sqrt{n_a^{\downarrow} n_b^{\downarrow}}\right) \left(1-\delta_{ab}\right) 
 \right] {\cal K}_{b}(\br)\,, \label{eq:fock}
\end{multline}
where 
\begin{eqnarray}
h(\br) & = & -\frac{\nabla^2}{2} + V_{ext}(\br)\,, \nonumber \\
{\cal J}_b(\br)  & = & \int d\brp \frac{\varphi_b^*(\brp) \: \varphi_b(\brp) }{|\br - \brp|}\,, \\
{\cal K}_b(\br) \varphi_a(\br)  & = & \int d\brp \frac{\varphi_b^*(\brp) \: \varphi_a(\brp) }{|\br - \brp|}  \varphi_b(\br) \,. \nonumber
\end{eqnarray}
A similar problem with an orbital-index dependent operator exists in the self-interaction-correction 
method\cite{svane,goda} of DFT and to a lesser extent in the ROHF method\cite{roothaan,hirao}.
Finally, the second of the subsidiary conditions, known as the N-representability constraint,
restricts the occupation numbers to lie between zero and one, $0 \leq n_a^{\sigma} \leq 1$
and guarantees that the 1-RDM corresponds to an $N$-body wave function. 

Our implementation  uses the GAMESS quantum chemistry program\cite{GAMESS} for the calculation
of the one-body and two-body integrals as well as for providing the Hartree-Fock solution which 
we choose as the starting point of our minimization procedure.  
The minimizations with respect to both the occupation numbers 
and the natural orbitals are performed with a conjugate gradient procedure. For the optimization
with respect to the orbitals, we adopted a procedure similar to the one described in Refs.~\cite{goda,cohen}.
We calculated the total energies of the first-row open-shell atoms, i.e. Li, B, C, N, O, and F. 
For all these elements we used the cc-PVQZ Gaussian-basis set\cite{ccPVQZ}. 

\begin{figure}
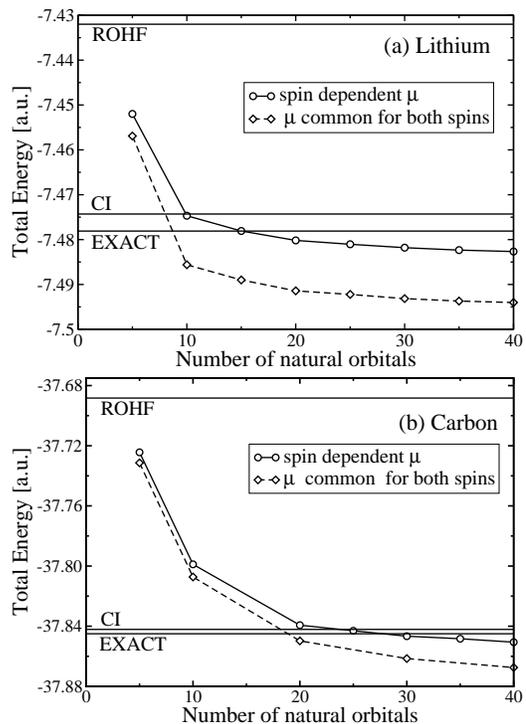

\begin{tabular}{c}
\includegraphics[width=0.38\textwidth,clip]{Li_Evsnnatorb.eps}\\ 
\includegraphics[width=0.38\textwidth,clip]{C_Evsnnatorb.eps} \\
\end{tabular}
\caption{\label{fig:e_vs_natorb} The convergence of the total energy with the number of natural orbitals
included in the minimization procedure for Lithium and Carbon atoms. The values of the ROHF method as well
as the CI and exact values are shown as horizontal lines.}
\end{figure}

In Figure~\ref{fig:e_vs_natorb}, we show the convergence of the total energy as a function 
of the number of natural orbitals included in the calculation. 
As it can be seen, 30-40 natural orbitals are typically required for full convergence. 
In Table~\ref{table:res}, we list the total energies for the open-shell systems as well as the He and 
Be closed-shell atoms for completeness. 
Comparing the total energies it becomes apparent that
forcing or relaxing the constraint of spin conservation is extremely important for open-shell
systems. The energy differences between the two minimization procedures are of the order of 10~mH with the 
spin-conserving constraint giving results much closer to the exact ones. Relaxing the constraint results 
in a charge transfer from one spin to the other which is of the order of 0.05-0.1 electrons. 
This charge transfer increases with the number of natural orbitals included in the minimization procedure 
which is the reason for the increase in the energy difference between the two procedures with the number 
of naturals orbitals seen in Fig.~\ref{fig:e_vs_natorb}. Interestingly, for the last two open-shell 
elements, i.e. O and F, where the 2p shell is more than half filled, the energy difference
between the two minimization procedures is smallest. 

It is clear from Fig.~\ref{fig:e_vs_natorb} and Table~\ref{table:res} that RDMFT offers a very 
good approximation for the correlation energy for all, closed and open-shell,
atomic systems we studied. Despite its simplicity, 
the functional we used produces results in very good agreement with state of the art, but 
computationally much more 
expensive, methods like CI.  However, the systematic trend of this particular
functional is to overestimate slightly the correlation energy for all the open-shell systems studied.
To give a fair credit to RDMFT one has to take into account that this is
one of the very first 1-RDM functionals that has been exploited in a minimization procedure, in
contrast to DFT functionals that have been heavily used and tuned for decades.

In Table~\ref{tab:mu}, we quote
the different values of the chemical potential for the majority and the minority spin as well as the
common value of it in the case of relaxing the constraint of spin conservation. Clearly, the values are
significantly different.
Interestingly, for B and C with the p-shell less than half filled, but also for N with half filled p-shell,
the majority $\mu^\uparrow$ is larger than the minority $\mu^\downarrow$. This is a consequence of the 
fact that ionization from the majority is energetically favorable. 
For the remaining two elements, i.e. O and F, with more than half filled p-shell this trend is opposite.

\begin{table}
\setlength{\tabcolsep}{0.2truecm}
\begin{tabular}{lcccc}
\hline\hline
 Atom &    \multicolumn{2}{c}{RDMFT}           &      QCI      &      Exact      \\
      & $\mu^\uparrow=\mu^\downarrow$ & $\mu^\downarrow\neq\mu^\uparrow$ &  &        \\ \hline
 He   &   \multicolumn{2}{c}{2.8978}           &    2.9049    &     2.9037      \\ 
 Li   &         7.4940     &        7.4827       &    7.4743    &     7.4781      \\ 
 Be   &   \multicolumn{2}{c}{14.6686}          &   14.6657    &    14.6674      \\ 
 B    &        24.6746     &       24.6616       &   24.6515    &    24.6539      \\ 
 C    &        37.8675     &       37.8506       &   37.8421    &    37.8450      \\ 
 N    &        54.6096     &       54.5965       &   54.5854    &    54.5893      \\
 O    &        75.0668     &       75.0671       &   75.0613    &    75.0670      \\
 F    &        99.6951     &       99.6952       &   99.727     &    99.734       \\
 \hline
 $\bar{\Delta}$ & 0.020 &    0.010 & 0.004 & 0.000 \\
 \hline\hline
\end{tabular}
\caption{\label{table:res} Absolute total energies for the first row atoms (in a.u.) with 
and without enforcing the spin conservation constraint. The QCI\cite{montgomery} and exact\cite{davidson}
values are also listed. $\bar{\Delta}$  is the mean absolute deviation from the exact values
(for open-shell systems only).}
\end{table}

\begin{table}
\setlength{\tabcolsep}{0.2truecm}
\begin{tabular}{lcc|cc}
\hline\hline
 & \multicolumn{2}{c|}{Forcing spin} &
\multicolumn{2}{|c}{Relaxing spin}  \\
 & \multicolumn{2}{c|}{conservation} &
\multicolumn{2}{|c}{conservation}  \\
 & $-\mu^\uparrow$ [a.u.] & $-\mu^\downarrow$ [a.u.] & $-\mu$ [a.u.] & $\Delta q$ [e] \\
\hline
He & 0.928 & 0.928 & 0.928   &    -     \\
Li & 0.191 & 2.541 & 0.186   &  0.012   \\
Be & 0.292 & 0.292 & 0.292   &    -     \\
B &  0.234 & 0.421 & 0.193   &  0.089   \\
C &  0.312 & 0.550 & 0.286   &  0.074   \\
N &  0.445 & 0.710 & 0.411   &  0.058   \\
O &  0.474 & 0.404 & 0.418   &  0.075   \\
F &  0.570 & 0.527 & 0.510   &  0.060   \\
\hline\hline
\end{tabular}
\caption{\label{tab:mu} The values of the spin-dependent chemical potential 
 $\mu^\uparrow$(the majority spin)  and $\mu^\downarrow$ in the case 
of enforcing the spin conservation constraint, and the common value of 
$\mu$ and the charge transfer $\Delta q$ in the case of relaxing that constraint.
} 
\end{table}

In summary, we have presented a systematic application of RDMFT to open-shell systems. We adopted 
a spin-restricted open-shell treatment and extended 1-RDM functionals to include spin-dependent
occupation numbers. This formalism has the advantage of allowing 
the prescription of a specific spin state. In particular,
we introduced a spin-dependent chemical potential in order to enforce 
conservation of the total spin in the minimization procedure,
in addition to the conservation of the total number of electrons.
We performed calculations for open-shell atoms using the appropriate extension of a standard
1-RDM functional\cite{mueller,UG}, both with and without the spin conservation
constraint. We conclude that the inclusion of this constraint is essential
for the proper treatment of open-shell systems. 
Our results for the total energies of the first row open-shell atoms 
are in very good agreement with state-of-the-art quantum-chemistry calculations. 
The presented formalism can be easily extended to any 1-RDM functional and 
therefore opens up the wide range of open-shell systems to the testing of new 1-RDM
functionals. 

\begin{acknowledgments}
We thank A. Zacarias for helpful suggestions.
This work was supported in part by the  Deutsche Forschungsgemeischaft within the program SPP 1145,
by the EXCITING Research and Training Network and by the NANOQUANTA Network of Excellence.
\end{acknowledgments}

\end{document}